\documentclass[twocolumn]{aastex62}

\usepackage{placeins}
\usepackage{natbib}
\usepackage{amsmath}
\usepackage{color}
\usepackage{enumerate}
\usepackage{graphicx}

\newcommand\msun{\rm M_{\odot}}
\newcommand\mdot  {\dot{M}}

\newcommand\kms{\rm \, km ~ s^{-1}}

\newcommand\beq{\begin{equation}}
\newcommand\eeq{\end{equation}}
\newcommand\be{\begin{equation}}
\newcommand\en{\end{equation}}
\newcommand\intalpha{\int_{t_{0,i}}^{t_{f,i}} dt \, \alpha_i}

\shorttitle{Gravity and mass functions} 
\shortauthors{Kuznetsova et al.}

\begin{document}

\title{The Role of Gravity in Producing Power-Law Mass Functions}

\correspondingauthor{Aleksandra Kuznetsova}
\email{kuza@umich.edu}

\author{Aleksandra Kuznetsova}
\affiliation{Department of Astronomy, University of Michigan, 1085 S. University Ave., Ann Arbor, MI 48109, USA} 
\author{ Lee Hartmann}
\affiliation{Department of Astronomy, University of Michigan, 1085 S. University Ave., Ann Arbor, MI 48109, USA} 
\author{Fabian Heitsch}
\affiliation{Department of Physics \& Astronomy, 3255 Phillips Hall, University of North Carolina, Chapel Hill,
North Carolina 27599, USA}
\author{ Javier Ballesteros-Paredes}
\affiliation{Instituto de Radioastronom\'ia y Astrof\'isica, Universidad Nacional Aut\'onoma de M\'exico, Apdo. Postal 72-3 (Xangari),\\ Morelia,
Michoc\'an 58089, M\'exico
}

\begin{abstract}
Numerical simulations of star formation have
found that a power-law mass function can develop at high masses.  In a previous paper, we employed isothermal
simulations which created large numbers of sinks over a large range in masses to show that the power law
exponent of the mass function, $dN/d\log M \propto M^{\Gamma}$, asymptotically and accurately approaches
$\Gamma = -1.$  Simple analytic models show that such
a power law can develop if the mass accretion rate
$\mdot \propto M^2$, as in Bondi-Hoyle accretion; however, the sink mass accretion
rates in the simulations show significant departures
from this relation.  In this paper we show that
the expected accretion rate dependence is more closely realized
provided the gravitating mass is taken to be the sum of the sink
mass and the mass in the near environment.
This reconciles the observed mass functions with
the accretion rate dependencies, and demonstrates
that power-law upper mass functions are essentially
the result of gravitational focusing, a mechanism present in, for example, the competitive accretion model. \end{abstract}


\section{Introduction}
Many analytic arguments and hydrodynamic simulations over the past few decades have been employed in attempts to understand the origin of the stellar initial mass function (IMF) \citep[see][for a recent review]{offner14}. In recent years, most such efforts have been numerical because of the complexity of the environments in which stars form. Many simulations now have managed to produce sink particle mass functions which are qualitatively similar to the IMF,
with a peak around $0.2 \msun$ and a width in mass consistent with observational estimates such as the \cite{chabrier2005} IMF 
(e.g., \citealt{bate12,krumholz12,lee18,haugbolle18}).

Because these simulations involve complex, turbulent regions with varying levels of detailed physics, the results
have been explained in qualitatively different ways.
Some authors emphasize the importance of density and velocity perturbations in local environments \citep[e.g.,][]{padoan02},
while others emphasize accretion from global scales \citep[``competitive accretion";][]{bonnell01a,bonnell01b,bate03}.  In addition,
although most investigators emphasize the importance of
thermal physics to establishing the peak of the IMF \citep{jappsen05,bate12}, this has been contested by \cite{haugbolle18}, who argue that isothermal MHD turbulence alone is sufficient.

While simulations differ in assumptions about equations of state, magnetic fields, stellar feedback, driven or decaying turbulence, etc., all ultimately rely on gravitational collapse to form stars.  This has led us to perform investigations which isolate the effects of gravity in star and star cluster formation with a minimum of additional complications.
In \cite{bp2015}  (BP15) we presented isothermal smooth-particle hydrodynamic (SPH) simulations with decaying turbulence and sink particle formation.  With this limited set of physics, we were able to produce enough
sinks to clearly show the development of a power-law distribution in masses.
While many other simulations have shown indications of power-law behavior at large masses \citep[e.g., Figure 5 of][]{jappsen05}, our simulations
covered a
sufficiently large dynamic range to allow a statistically robust estimate of the IMF slope of   $dN/d\log M = M^{\Gamma}$, with
$\Gamma = -1 \pm 0.1$.

To further simplify the physics involved and focus on
the effects of gravity, in \cite{kuznetsova2017} we showed that pure $N$-body simulations with an initial
set of velocity fluctuations also produce power-law mass functions of similar slopes.  
Since the classical Salpeter value of $\Gamma = -1.35$ for the stellar IMF is only slightly steeper, and as the young cluster mass function is estimated to be $\Gamma \sim -1$ \citep{lada03,fall12}, we have argued that
gravitational physics is dominant in collecting
the mass that forms both high-mass stars and star clusters.

The robust value of 
$\Gamma = -1$ found in these simulations led us to suggest that
something like Bondi-Hoyle-Littleton (BHL) accretion \citep{edgar2004} was occurring.
\cite{zinnecker1982} (Z82) long ago showed that this
mass function slope could be produced by BHL accretion,
with a dependence of mass accretion rate on accreting mass of $\mdot \propto M^2$, as long as
enough mass is added to the original distribution of ``seed" masses.
In this view, the Salpeter slope is steeper
because mass growth is often terminated before
the fully asymptotic limit can be realized.

However,
the standard BHL formula assumes a non-self-gravitating environment of constant density and relative velocity that is not realized in simulations (or in real star-forming regions). 
Furthermore, the Z82 model assumes sink accretion
rates $dM/dt \propto M^2$ which is
only roughly (at best) exhibited in simulations \citep{hsu2010}, partly due to variations in environmental densities and velocities (BP15). These departures led
\cite{maschberger14} (M14) to 
argue that BHL accretion was not occurring in their simulations, even though they obtained a Salpeter-like upper IMF slope.

In an attempt to further clarify how gravitationally-driven accretion proceeds in
simulations of star formation, in this paper 
we present additional numerical experiments with sink
formation in isothermal clouds with decaying turbulence.
We show that to understand the
accretion process it is necessary to consider the self-gravity of the near environment of
sinks, not just that of the sink itself. 
We conclude that 

gravitational focusing is an effective process for power-law mass functions with $\Gamma \sim -1$.

\section{Numerical experiments}

\subsection{Basic Assumptions and Sink Implementation}

\par We use the Eulerian fixed grid code \emph{Athena} \citep{stoneetal2008} to simulate the collapse of a molecular cloud by self-gravity. We solve the system of equations
\begin{eqnarray}
\label{eq:mass}
\frac{\partial\rho}{\partial t} + \nabla \cdot (\rho \mathbf{v}) = 0\\
\label{eq:mom}
\frac{\partial \rho \mathbf{v}}{\partial t} + \nabla \cdot (\rho\mathbf{vv} + P) = -\rho \nabla \Phi\\
\label{eq:poisson}
\nabla \Phi = 4\pi G \rho
\end{eqnarray}
with periodic boundary conditions.
To improve the time order of the scheme, especially in the context of sink dynamics and accretion, we implemented 
a RK3 integrator \citep{GottliebShu1998}, which advances the fluid equations (eqs. \ref{eq:mass} and \ref{eq:mom}) at third order in time.
We further adopt an isothermal equation of state such that $P=c_s^2 \rho$ for simplicity, which is a reasonable approximation for low-mass star-forming regions on the scales we study. The Poisson equation (eq.~\ref{eq:poisson}) is solved every RK3 substep, using the FFT solver that comes with the stock version of Athena.

We adopt a sink and near-environment (``patch") geometry similar to those in \cite{bleulerteyssier2014} and 
\cite{gongostriker2013}.  The motivation for the sink implementation is twofold; to enable modeling of the dense peaks that occur during gravitational collapse and to study the gas properties of the sink environs.
The \cite{truelove97} criterion sets the maximum ratio of the cell size $\Delta x$ to the Jeans length allowed to avoid artificial fragmentation; in the isothermal case, this corresponds to a maximum density 
\begin{equation}
  \rho_{\rm t}=\frac{\pi}{16}\frac{c_s^2}{G\Delta x^2},
\end{equation}
with the sound speed $c_s$ and the (fixed) cell size $\Delta x$.
Sink-patches can be created when the density peak across a group of cells
exceeds $\rho_{\rm t}$,  
the flow is converging such that $\nabla \cdot v < 0$ in the patch, and the patch gas is bound, $E_K + E_G < 0$,
 where $E_K$ is the sum of both bulk motion and thermal energy. The gravitational energy $E_G$ is calculated from the gravitational potential, rather than from the standard equation $3GM^2/5R$, since there can be significant misinterpretation on whether the core is bound or unbound \citep[see][for a discussion]{ballesterosetal18}.

Since the sink patches do not necessarily align with the underlying grid, their content is recalculated at every timestep. Patches are allowed to overlap, but sinks are made to merge when one sink enters another sink's patch. Sinks are initialized with a token mass ($10^{-4} \rho \Delta x^3$) proportional to the local density, and they gain the rest of their mass by accretion from the patch, the reservoir for the sink. The accretion rate onto the sink is dictated by the (net positive) mass flux into the patch and regulated by the ratio of the mean density $\bar{\rho}$ over the threshold density dictated by the Truelove criterion, $\rho_t$ (Eq \ref{eq:bt1}, see \citealp{bleulerteyssier2014} for details), 
\begin{equation}
    \dot{M}_s = \left(1+ \eta\log \frac{\bar{\rho}}{\rho_{\rm thr}}\right) \int \nabla \cdot (\rho(\mathbf{v-v}_{\rm sink})) dV,
\label{eq:bt1}
\end{equation}
where $\eta$ is an adjustable parameter.  We have experimented with both $\eta = 0.1$ and $\eta = 1$; the results are not sensitive to the particular choice.
The resulting sink mass accretion rate $\dot{M}_{\rm sink}$ and the local patch densities $\rho$ determine how much mass $\Delta m$ is removed from each cell in the patch by
\begin{equation}
  \Delta m = \Delta t \frac{\dot{M}_{\rm sink}}{n_{\rm cells}}\frac{\rho}{\bar{\rho}},
 \label{eq:bt2}
\end{equation}
where $n_{\rm cells}$ is the number of cells in the sink patch.
Equations \ref{eq:bt1} and \ref{eq:bt2} regulate the flow of
mass onto the sink so as to avoid large jumps in density or velocity. 
Since mass is removed instantaneously across the patch (but spatially varying, depending on the local density, see eq.~\ref{eq:bt2}), the size of the patch is the relevant resolution element.

The number of patch cells is tuned by the accretion radius parameter $r_{\rm acc}$, which determines the number of cells, in addition of the center cell that contains the sink, that make up the radial extent of the patch. The simulations presented here have an $r_{\rm acc} = 2$, such that the resulting patches contain $5^3$ cells in total.  Therefore, the linear sink patch size is only slightly larger than the stencil size used for reconstructing the hydrodynamic quantities (3 cells), which determines the actual numerical resolution.
Sink-patch data is output as the three dimensional sink velocity, sink position, and the conserved variables in the patch cells (and an additional set of boundary cells).

The gravitational interaction between sinks and between sinks and gas is calculated via the Fourier gravity solver that comes with Athena. To that purpose, the sink mass is distributed on the density grid before the gravity-solve, via the triangular-shaped-cloud scheme using three support points \citep[][see also \citealp{gongostriker2013}]{hockneyeastwood1981}. Note that this interpolation occurs on the underlying grid, not within the sink patch. Therefore, the sink is not necessarily located at the center of a cell. The standard kernel used in the Fourier solver implies a gravitational softening of one cell. Though this method is not capable of accurately calculating close sink encounters, this is irrelevant for our purposes, since (a) sinks merge when entering each other patches, and (b) the run times are too short for sinks to orbit around each other.

\begin{figure*}[tbh]
    \centering
    \includegraphics[width=0.95\textwidth]{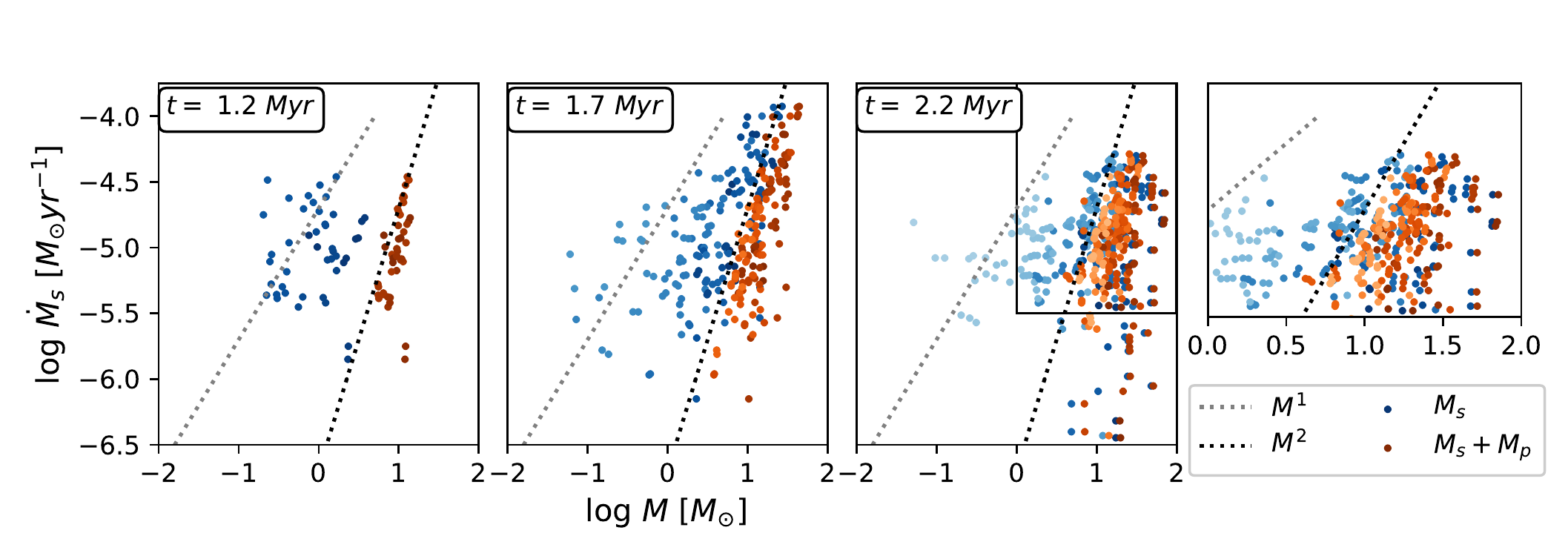}
    \caption{Accretion rate onto the sink $\dot{M}_s$ vs sink masses alone $M_s$  (blue) and total mass enclosed within the patch boundary, a sum of the sink mass and patch gas mass $M_s + M_p$ (orange) combining several outputs around 1.2 Myr, 1.7 Myr, and 2.2 Myr in the simulation. Lighter shades of blue and orange are sinks that have formed later in the simulation. The gray dotted line goes as $M^1$ and the black dotted line as $M^2$. The inset is a zoom of the x dimension to show the scatter in mass and corresponds to the area of the rectangle drawn in the third panel. } 
    \label{fig:mdotm}
\end{figure*}

\subsection{Parameters and Initial Conditions}
\par All the runs used in this paper are of a 4 pc radius spherical, constant density cloud in a 20 pc box. The initial density of the cloud is $\rho_c = 1.5 \times 10^{-21} \mathrm{g cm^{-3}} $, which is 100 times the ambient density in the box, giving the cloud an initial free-fall time $t_{\mathrm{ff}} = 1.7$ Myr.  
The simulations are isothermal with a temperature of 14 K, 
assuming molecular gas with $\mu = 2.4$.
We set up a decaying supersonic turbulent power spectrum $P(k) \propto k^{-4} dk$, initially at Mach 8, which leads to an initial velocity dispersion of $\sim 2 \kms$ typical of molecular clouds. These supersonic motions dissipate rapidly and the velocity dispersion of the cloud is driven by gravity as the cloud collapses, becoming roughly virial as observed in star-forming clouds\citep[see ][]{BP+11, ballesterosetal18}.
 
We ran simulations with $N_{\rm cell} = 256^3, 512^3, 1024^3$, for which the cell size $\Delta x = 0.08,$ $0.04,$ $0.02$ pc and the patch radii are $r_p = 0.2,$ $0.1,$ $0.05$ pc, respectively. We present our fiducial runs at $512^3$ in the following section, and discuss results for the $1024^3$ run in the Appendix. Because these simulations are isothermal, the smallest resolvable structures scale with the grid, and
therefore details of the fluid dynamics will not converge with increased resolution. However, because isothermal simulations can be rescaled \citep{hsu2010},\footnote{ If the size scale is changed by a factor
of $D$ and the total mass is similarly scaled by $D$, the density changes 
by $D^{-2}$; thus for the same gas temperature the ratio of the sound crossing
time to the free-fall time is the same, and the number of initial Jeans 
masses in the cloud also remains fixed.} runs with differing
grid sizes essentially test whether changing the dynamic range of the simulation affects the overall results.  Therefore, while we cannot expect convergence in the details of density and velocity fields, statistical measures such as mass functions can "approach convergence" when increasing the resolution \citep[e.g.][]{haugbolle18}.

\begin{figure*}
    \centering
    \includegraphics[width=0.95\textwidth]{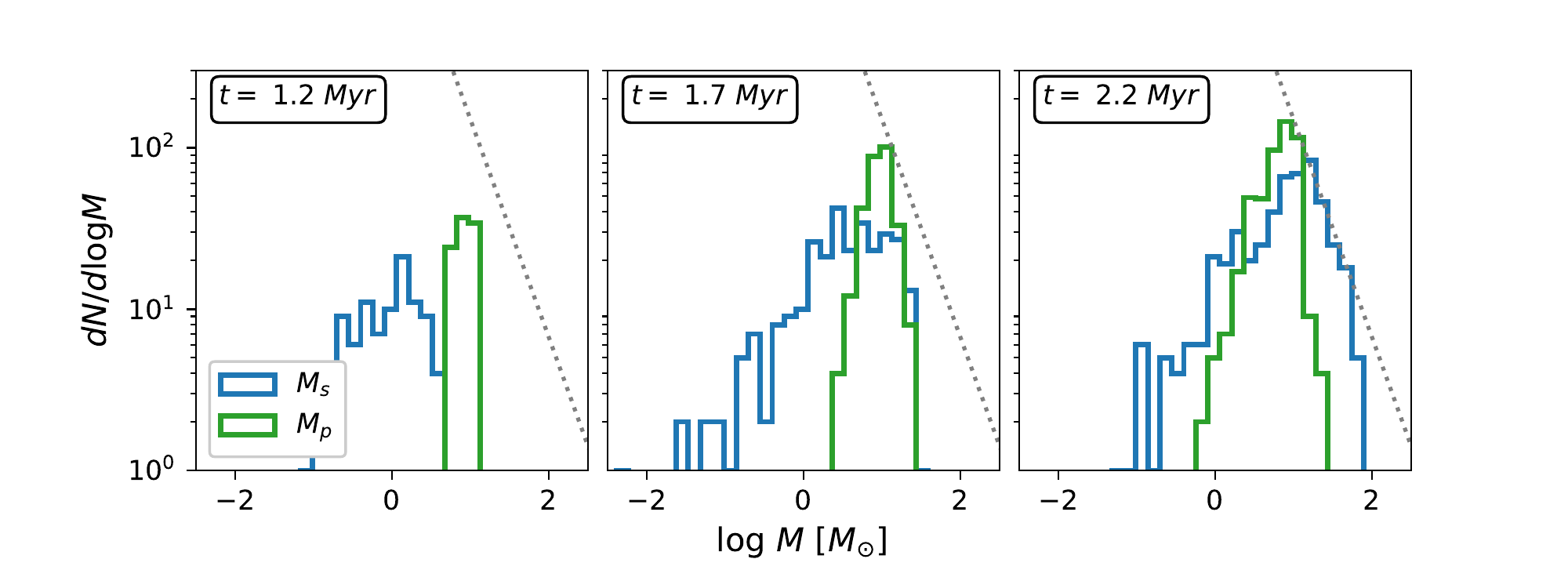}
    \caption{Mass functions for the sink mass $M_s$ (blue) and the patch gas mass $M_p$ (green) at 1.2, 1.7, and 2.2 Myr into the simulation. The gray dotted line is the Salpeter slope.}
    \label{fig:mf}
\end{figure*}

\section{Results}

 In Figure \ref{fig:mdotm} we plot the relationship between the sink accretion rate $\dot{M}_s$, the mass of the sinks $M_s$, and in contrast, the total mass enclosed within the patches $M_s + M_p$ at three different times in the simulation $t = 1.2,\ 1.7,$ and $2.2 {\, \rm Myr} $. This leads to an important insight: the mass-squared dependence is seen much more
clearly when using
the total mass $M_s + M_p$. At 1.2 Myr,
when the first sinks are forming, the accretion rate is tightly 
correlated with the 
$\mdot_s \propto (M_s + M_p)^2$ dependence, and the correlation is still strong at 1.7 Myr.
Near the end of the simulation at 2.2 Myr, the scatter in the accretion vs. total mass relation increases significantly, though the overall mass-squared dependence remains. It is likely that the increased scatter
is due to the depletion of the environment as mass accretes onto
sinks (BP15), especially for the very massive sinks with low accretion rates; these have strongly depleted their surrounding gas.

In contrast, the accretion-sink mass relation exhibits much larger scatter and only approaches $\mdot_s \propto M_s^2$ at late times.  The reason why the dependence
of the accretion rate on sink mass is so
different is that the sink masses constitute only a
fraction of the total gravitating mass throughout
most of the simulation.
As shown in Figure \ref{fig:mdotm}, and emphasized by the mass functions shown in 
Figure \ref{fig:mf}, sink masses can be an order
of magnitude smaller than patch masses
at early times. It is only
near the end of the simulation that many
sink masses become comparable to or
larger than patch masses, at which point the
$\mdot_s - M_s$ relation approaches the
mass-squared dependence.

Thus, the main 
objection to attributing the growth of a power-law mass function to BHL-type accretion, that the $\mdot \propto M_s^2$ dependence is weakly if at all present (e.g., M14, BP15), is shown to be the result of
misidentifying the gravitating mass as only
that of the sink.

\begin{figure*}
    \centering
    \includegraphics[width=0.9\textwidth]{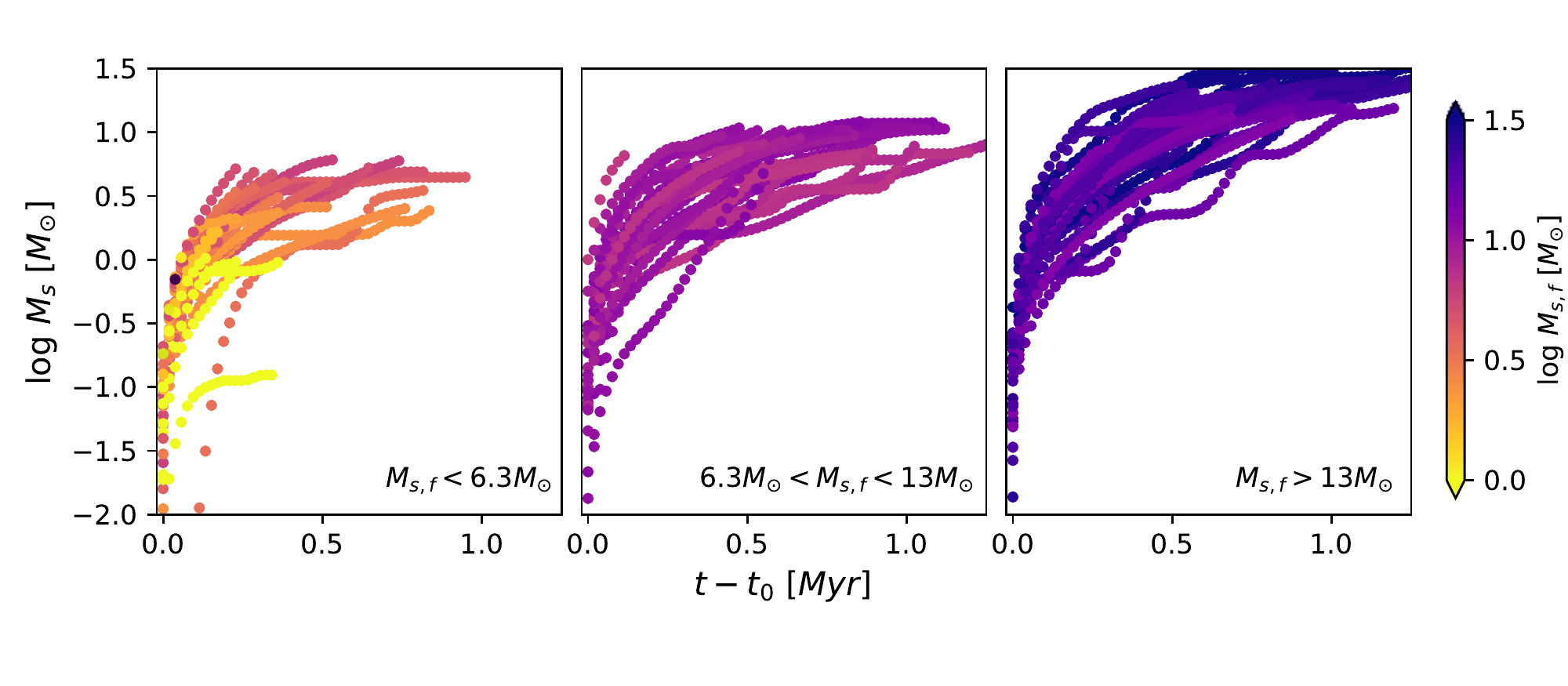}
    \caption{The sink mass growth over time, zeroed at the formation time for each sink $t_0$, shown for three populations binned and color-coded by the final sink mass. From left to right; sinks with a final mass $M_{s,f} < 6.3 M_{\odot}$, $6.3 M_{\odot} < M_{s,f} < 13 M_{\odot}$, $M_{s,f} > 13 M_{\odot}$. All sinks grow rapidly within the first few 0.1 Myr, but turn over at different rates.}
    \label{fig:mvt}
\end{figure*}

 One question is whether our prescription for sink accretion differs  sufficiently from that of the SPH simulations of M14 and BP15, and from some other implementations in grid codes,  to strongly affect our basic conclusion.  We find that in the simulations shown, with $\eta = 0.1$ (equation \ref{eq:bt1}), the accretion rate onto the sink is generally $\sim 90 - 95$\% of the accretion rate into the patch, so Figure \ref{fig:mdotm} would be essentially unchanged if the patch accretion rate were used instead of the sink accretion rate; we use the latter for comparison to the literature. (Note that the density and velocity distributions within the patch must be uncertain due to limited resolution of the hydrodynamics, such that more complex prescriptions are not clearly warranted.)

\begin{figure*}
    \centering
    \includegraphics[width = 0.9\textwidth]{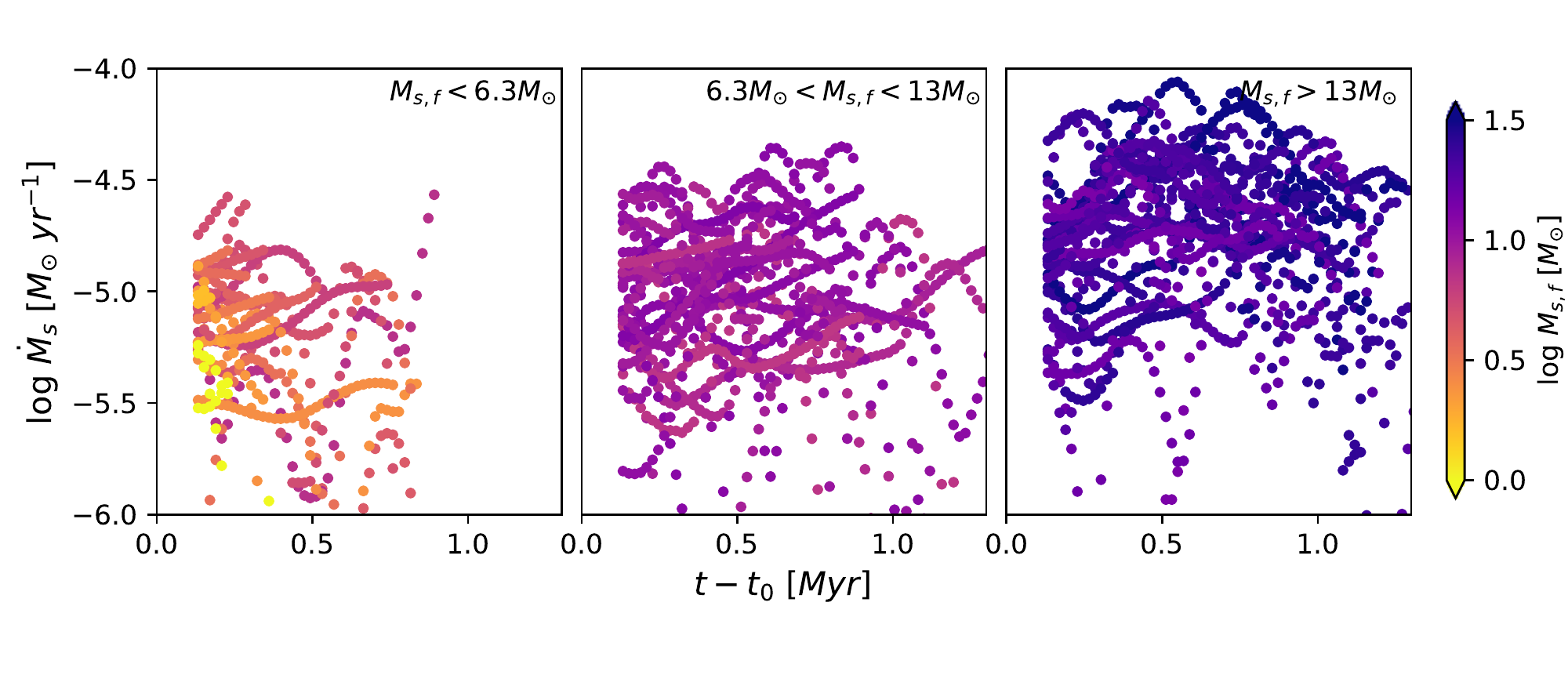}
    \caption{Time averaged sink accretion rate $\dot{M}_s$ as a function of the time since formation $t-t_0$, binned and color coded by final sink masses for the same sinks as in Figure \ref{fig:mvt}. Each panel represents a different final sink mass range. From left to right: lower mass sinks - $M_s < 6 M_{\odot}$, intermediate mass sinks - $ 6 M_{\odot} < M_s < 13 M_{\odot}$, and highest mass sinks - $M_s > 13 M_{\odot}$. Individual sink mass growth rates are highly variable, even when averaged across 0.1 Myr windows.}
    \label{fig:mdotvt}
\end{figure*}

\begin{figure}
    \centering
    \includegraphics[width=0.9\columnwidth]{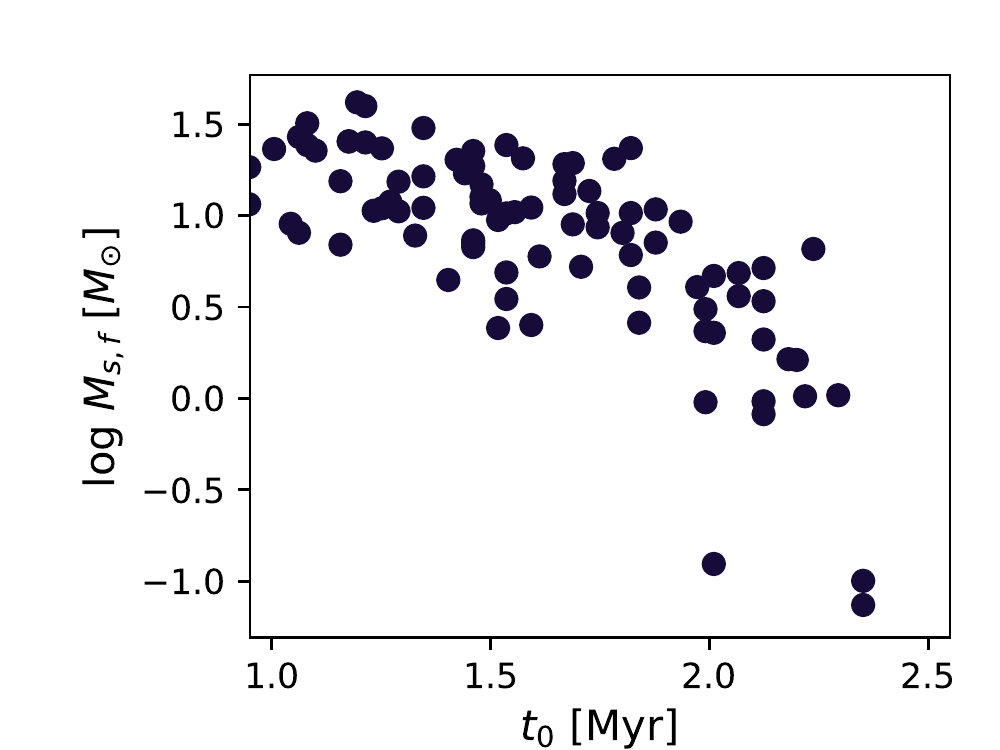}
    \caption{Final mass of the sinks in the simulation vs their formation time in the simulation. 
    }
    \label{fig:tform}
\end{figure}

More generally, our results for the accretion rates onto sinks as a function of time show the characteristic ``banana-shaped'' shape seen the SPH simulations (e.g., Figure 1 in M14 and Figure 14 in BP15), the large fluctuations in accretion rates with time (Figure \ref{fig:mdotvt} as in M14 (their Figures 1 and 2), and a correlation of final sink masses with initial times of formation 
(Figure \ref{fig:tform}; cf. Figure 12 of M14 and Figure 15 of BP15). 
Thus, there is no indication from the behavior of $dM_s/dt$ for individual sinks for a qualitative difference in behavior from M14 and BP15, despite the differences in
the algorithms for sink accretion.

In addition, as in other simulations with
sink formation in a turbulent environment
(e.g., Figures 1 and 2 in M14),
 ``instantaneous'' accretion rates show
frequent fluctuations of an order of magnitude or more over timescales of
$\sim 0.1 - 0.3$~Myr (Figure \ref{fig:mdotvt}).  In our case, these fluctuations are
related to the clumpy and moving filamentary structure of the gas in the sink environments
 \citep{Smith_2010}.

 As shown in the Appendix, increasing the resolution to $1024^3$ results in the formation of a larger number of sinks with smaller masses; however, the main conclusion of this paper, that the $dM_s/dt$ tracks
$(M_s + M_p)^2$ much more closely than $M_s^2$, is unchanged.

These simulations only begin to provide
an indication of a power-law upper mass function
at 2.2 Myr (Figure \ref{fig:mf}).
This is consistent
with the argument from the simple BHL accretion model of Z82, in which the
power-law distribution only develops once the masses have grown well beyond their initial values.  In our case, the relevant
initial mass distribution is not that of the sinks but that of the patches, as the patches set our effective
resolution limit for the dynamics.  With
initial patch masses $\sim 5-10$ $\msun$ (left panel of Figure \ref{fig:mf}), mass growth to $\sim 10^2 \msun$ is required to produce
sufficient dynamic range to indicate a
power-law behavior.  (We note that the power-law distribution is more apparent in the higher-resolution run; see Figure A.2).

\section{Discussion}
\label{sec:discussion}

The essence of the BHL accretion formula is
the concept of accretion from material
passing within a gravitational capture radius
$r_g \sim G M v^{-2}$, where in the simplest
model $v$ is flow past the point mass.  Then
\begin{equation}
\frac{dM}{dt} = \alpha M^2\,,
\label{eq:mdotalpha}
\end{equation}
where $\alpha \approx 4 \pi G^2 \rho v^{-3}$,
where $v^2 = v_b^2 + c_s^2$, $v_b$ and $c_s$ being the bulk flow velocity and sound speed, respectively
\citep{edgar2004}.
Z82 then showed
that if $\alpha =$~constant, then the mass of
an individual object grows with time as
\begin{equation}
    M(t) = \frac{M_0}{1 - \alpha M_0 t}\,
    \label{eq:moft}
\end{equation}
where $M_0$ is the initial mass at $t = 0$. 

\begin{figure}
    \centering
    \includegraphics[width=0.9\columnwidth]{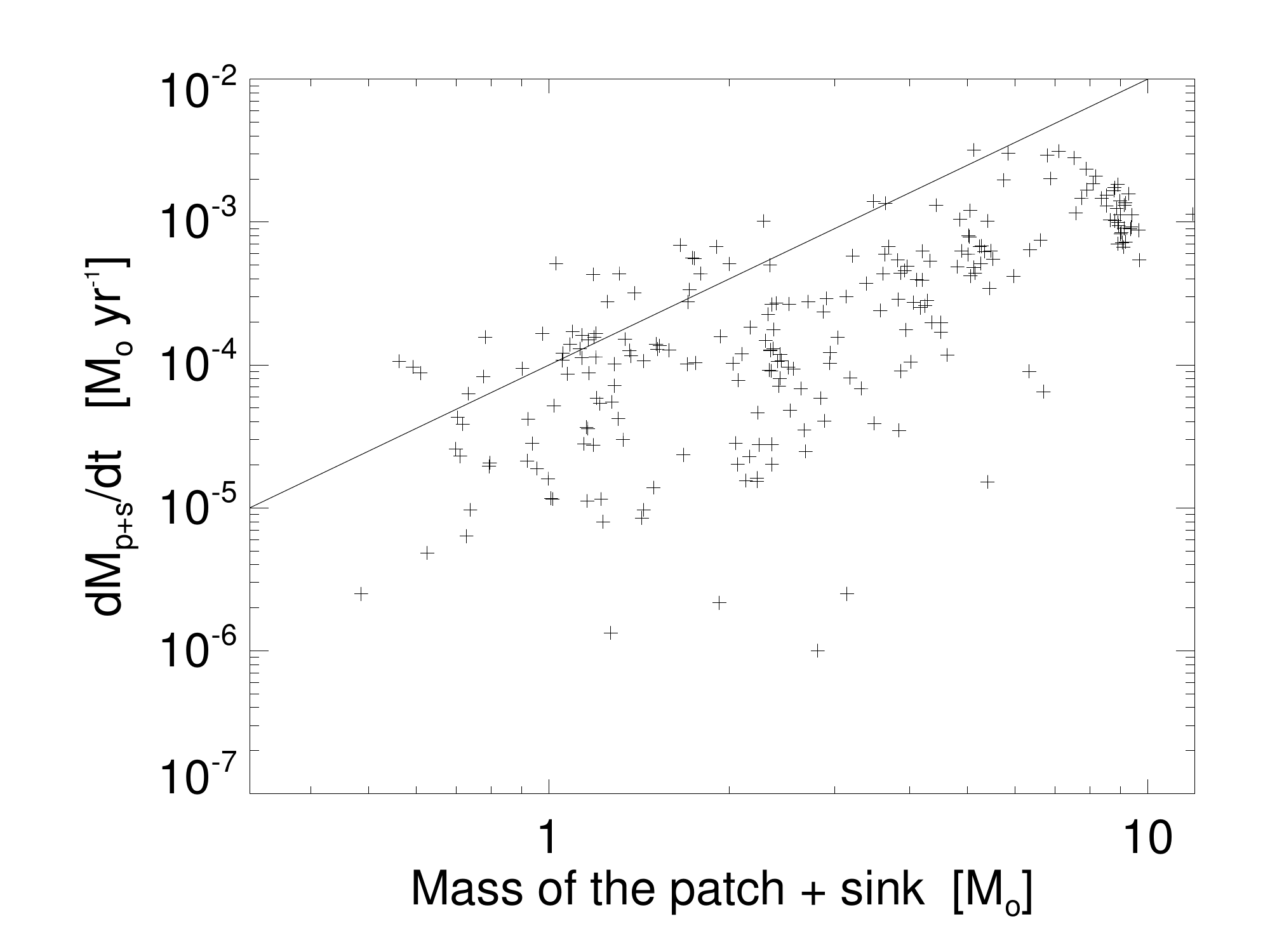}
    \caption{Accretion rates for run 22 of BP15 at $t = 2.5 \times 10^4$ yr, plotted as a function of the total mass within 6 $R_{\rm outer} = 1.08 \times 10^{-3}$~pc, where $R_{\rm outer}$ is the external radius where the physical properties are evaluated to determine whether an sph particle will be accreted or not (see BP15).  The $M^2$ relation is clearer than using simply the sink mass (compare with Figure 2 of BP).}
    \label{fig:mdotgadget}
\end{figure}

M14 argued that BHL-type accretion was not occurring in their
simulations, based on the departure of $\mdot$ from a $M_s^2$ relation. However, we have shown in our simulations that the mass-squared dependence {\em does} occur for $\mdot$ as a function of the total mass $M_s + M_p$, though with some scatter. Furthermore, we can check this result by revisiting the data from BP15.
In Figure \ref{fig:mdotgadget} we show a
reanalysis of the BP15 results now
plotting the mass accretion rate vs.
the sum of the sink mass plus the mass
of the near environment.
As in the \emph{Athena} simulations, the resulting distribution much more clearly exhibits 
the mass-squared dependence (compare with 
Figure 2 in BP15), again with
significant scatter.
The recognition of the importance
of mass in the near-sink environment thus demonstrates how the
power-law distribution $\Gamma \rightarrow -1$ seen in the
BP15 simulations arose.

Thus, the gravitating mass that should be considered in reference to equation \ref{eq:mdotalpha} is the total in the near environment, not just that of the sink.
This is especially important at early times, when the sink mass is much lower than the patch mass.  As seen in Figure \ref{fig:mdotm},
when the sink mass grows to be comparable to or larger than that of the patch mass, the mass accretion rate approaches $M_s^2$ as expected.

The fast initial rise in $\mdot$
as a function of the sink mass is qualitatively different than predicted by equation (\ref{eq:moft}).  As argued by
M14 and BP15, this phase is analogous to the
formation of a central protostar surrounded
by a massive envelope,
as seen for example in the collapse of Bonnor-Ebert spheres or clouds of similar structure,
which show an initial spike of very rapid
central accretion \citep{foster1993,henriksen97}.
However, at the same time as 
the central protostar (sink) is
undergoing this rapid phase of accretion, 
the protostellar envelope can be accreting at the mass-squared accretion rate, as shown in Figure \ref{fig:mdotm}.

Our patch masses, with their arbitrary size scale, are only a crude proxy for the gravitating mass responsible for accreting
gas from the general environment.
Indeed, it is not clear what prescription one should use
to distinguish between ``protostellar core/envelope'' and ``external environment'' in a
continuous medium.The size scale over which the accretion rate is measured naturally affects the value of $\dot{M}_s$, but the mass-squared dependence persists over a range of tested patch sizes  - from 0.05 pc to 0.25 pc. 
The much stronger correlation of
accretion rate with total mass, and its much closer approach to a mass squared dependence, adequately illustrates the need to consider scales well beyond
that of the sink to see the  mass-squared relation.

\begin{figure}[ht]
\centering
\includegraphics[width=0.45\textwidth]{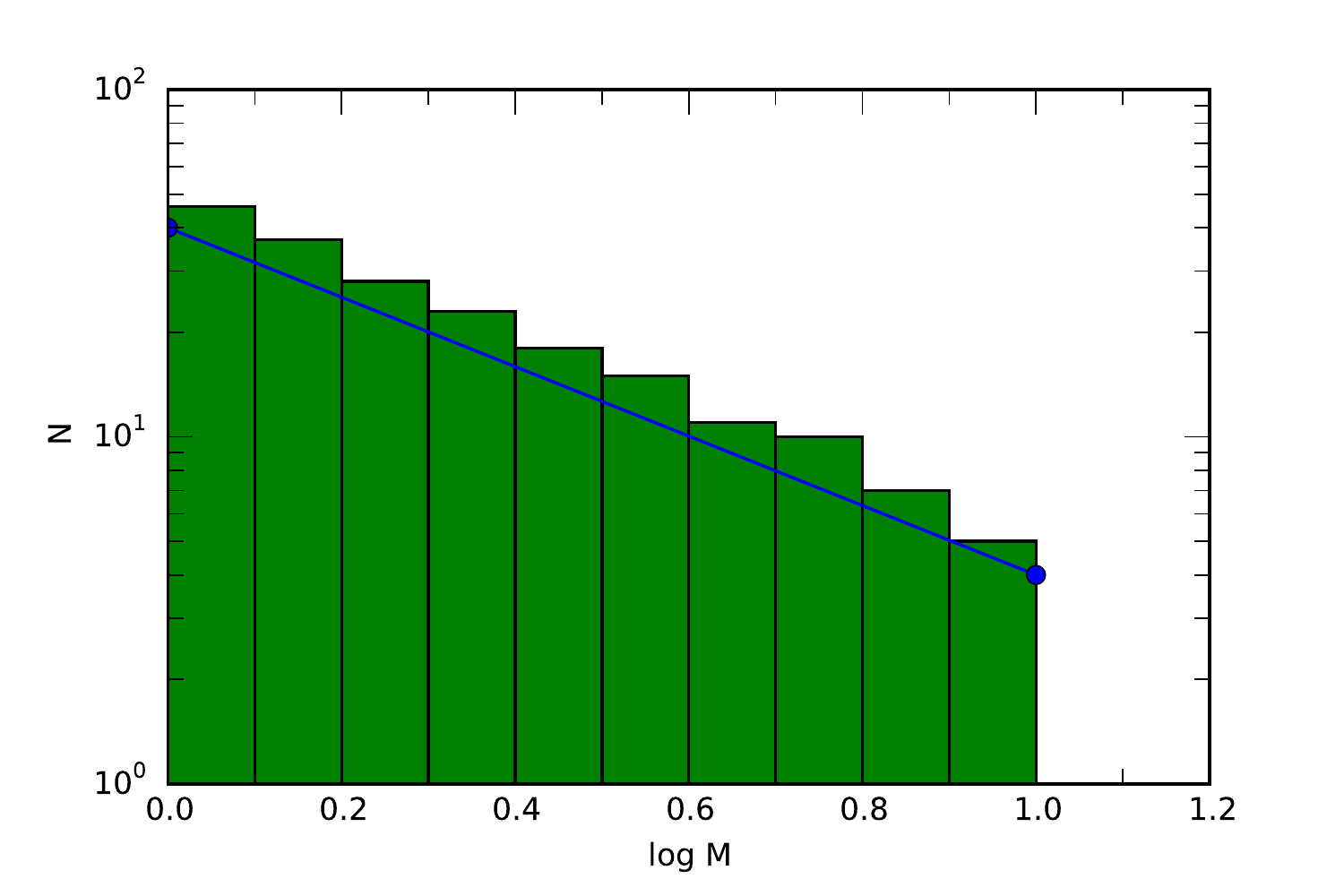}
\caption{Histogram of the mass distribution using
equation \ref{eq:inversem} for a population of 200 objects with an initial starting mass of unity with uniformly spaced values of
$\langle \alpha t \rangle_i$ between 0 and 0.9 (see text).  The straight line is not a fit,
but demonstrates a slope of $\Gamma = -1$.}
\label{fig:toyimfinv}
\end{figure}

One common feature of simulations with
sink formation in a turbulent environment is
that 
(e.g., Figures 1 and 2 in M14),
 ``instantaneous'' accretion rates show
frequent fluctuations of an order of magnitude or more over timescales of
$\sim 0.1 - 0.3$~Myr (Figure \ref{fig:mdotvt}; see also 
Figures 1 and 2 in M14).  These fluctuations are
related to the clumpy and moving filamentary structure of the gas in the sink environments
 \citep{Smith_2010}.  The result is a scatter in $\mdot$ that tends to obscure the overall mass dependence.  Furthermore,
the increasing scatter in the $\mdot\ {\rm vs.}\ (M_s + M_p)$ relation as the simulation
proceeds is arguably due to depletion of the environment as the sinks accrete mass.  This of course is
inconsistent with the Z82 model (cf. equation \ref{eq:moft}; Figure \ref{fig:mdotvt}). 

M14 argued any theory of the mass function
must take the long- and short-timescale variations of accretion into account, as well as the difference in accretion times. However, this is not necessarily the case.
Assume accretion is given by equation \ref{eq:mdotalpha} holds for each mass $M_i$;
then integration yields
\begin{equation}
\frac{1}{M_i} = \frac{1}{M_{i,0}} - \langle \alpha t \rangle_i\,,
\label{eq:inversem}
\end{equation}
where $M_{i,0}$ is the initial mass and
$\langle \alpha t \rangle_i \equiv \intalpha$.
Starting from equation \ref{eq:inversem},
Z82 showed that for constant $\langle \alpha t \rangle_i = \alpha t$,
an initial distribution of masses $M_{i,0}$ could grow to produce
a power-law distribution of final masses $M_i$ with $\Gamma \rightarrow -1$
asymptotically.  However, it is clear
that alternatively one can fix $M_{i,0}$ and still produce the power law distribution
of $M_i$ by adopting a suitable variation in $\langle \alpha t \rangle_i$.

To provide a simple demonstration of how a range in accretion rates and times can produce a power law mass distribution, we use equation \ref{eq:inversem} to calculate the final masses for 200 objects, all with the same initial mass $M_{0,i} =1$, and with a uniform spacing
in $\langle \alpha t \rangle_i$ between 0 and 0.9.
The
resulting histogram of the $M_i$ distribution is shown in
Figure \ref{fig:toyimfinv}, demonstrating the emergence of a $\Gamma \sim -1$
power law (straight line).

Thus, variation in $\langle \alpha t \rangle$, due either to varying $\alpha(t)$ or $t_f - t_0$, or both,
need not prevent the development of the
upper mass power law with $\Gamma \rightarrow -1$.
In particular, simulations with $t_f =$~constant but differing $t_{0,i}$ can be accommodated with a suitable distribution of starting times; all that matters is the total mass accretion at the end of the calculation.
As the final mass will tend to be larger for those objects that form first, as seen in 
Figure \ref{fig:tform}. Of course, not every possible distribution of $\intalpha$ and $M_{0,i}$ will produce a clear asympotic power law mass distribution in $M_i$; further exploration of various forms may be instructive.

\begin{figure*}
    \centering
    \includegraphics[clip=true, trim=0 2.25in 0 0 ,width=0.86\paperwidth,]{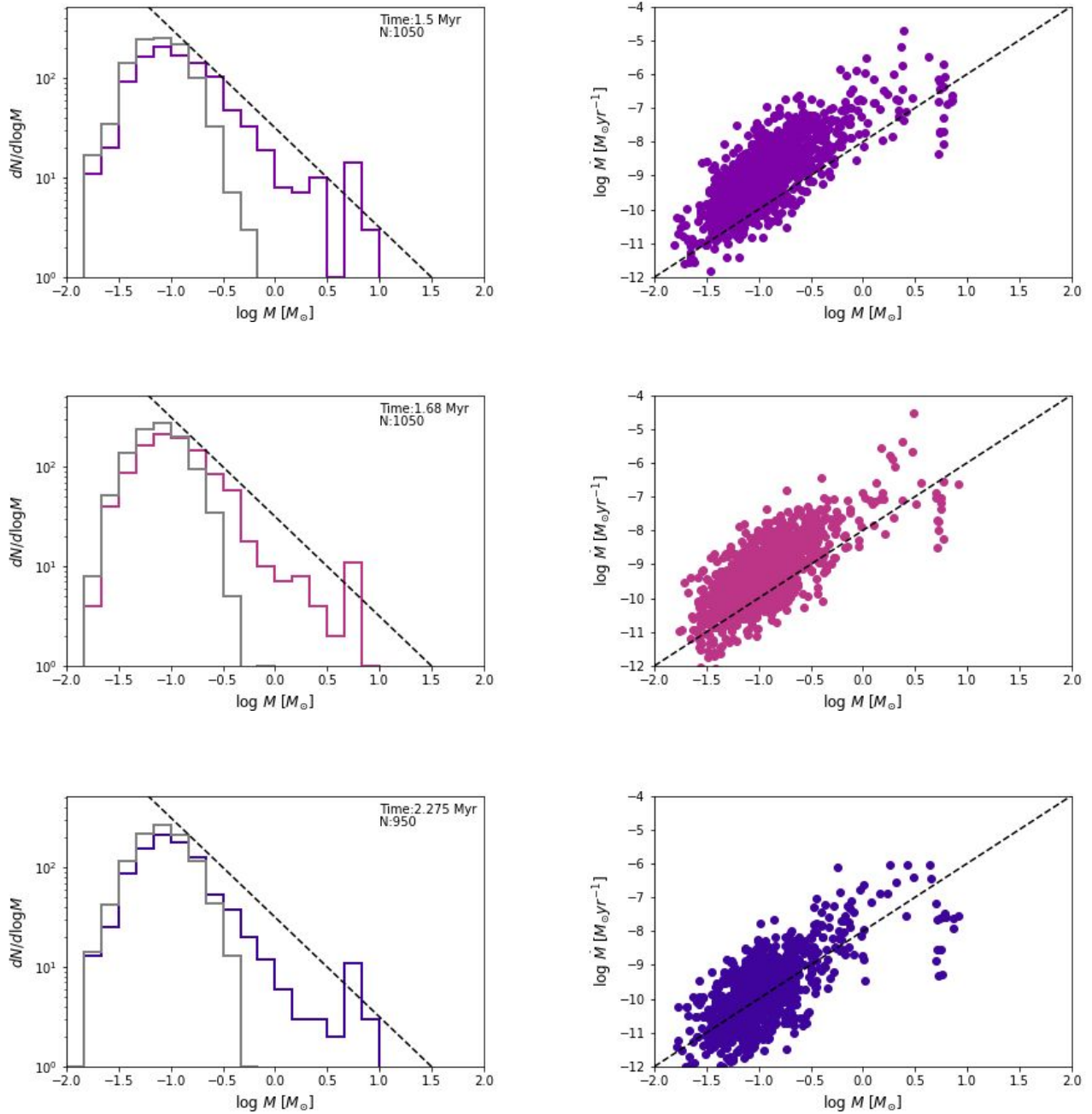}
    \caption{Mass functions $\xi(M)$ (Left) and $dM/dt$ vs $M$ (Right) for three types of simulation prescriptions. The initial $\xi(M_0)$ distribution sampled is shown as a light gray histogram. The dashed line on the mass function plot has a slope of -1 in dN/dlogM space. The dashed line on the $dM/dt$ plot has a slope of 2. From top to bottom: a constant log-normal $\alpha$ distribution shown at $t= 1.5$ Myr - scatter originates from random sampling and spread in $\alpha$. Middle row: A log-normal distribution of $\alpha$ that shifts to lower values over time shown at $t = 1.68$ Myr, where the primary difference is that it takes slightly longer to develop the power law tail. Bottom row: The same decreasing distribution of $\alpha$, with a spread in star formation - that is the $\xi(M_0)$ is sampled at a specified star formation rate, forming 50 stars every 0.125 Myr, shown here at $t=2.275$ Myr. Note that the distribution of $dM/dt$ appears to have a different slope with regards to the mass, despite having the same $\dot{M} \propto M^2$  accretion prescription. }
    \label{fig:toybhl}
\end{figure*}

The actual simulation behavior is of course more complicated than in the above example. We therefore conducted some simple numerical simulations in which we evolve a population of stars with distribution of masses $\xi(M_0)$ through time as they accrete according to a BHL-like prescription  \citep[][]{akuznetsova_18}\footnote{Codebase: \url{https://github.com/akuznetsova/BHL} }. At each timestep, each star's current mass $M_i$ is evolved to it's mass at the next timestep $M_{i+1}$ with the equation:

\begin{equation}
    M_{i+1} = M_i + \alpha_i M_i^2 dt
\end{equation}
where $\alpha_i$ is drawn randomly from a log-normal distribution. In Figure \ref{fig:toybhl}, we show $\xi(M)$ and $dM/dt$ vs.\ $M$ for three different cases all with a similar $\xi(M_0)$ drawn from a log-normal distribution centered at $\mu= 0.08 M_{\odot}$ with a spread of $\sigma_{\log M} = 0.25 $, where $N_{\rm total} = 1050$, and all with the same starting $\alpha_0$ distribution. For all cases, when  masses reach a threshold mass of 8 $M_{\odot}$, the $\alpha$ value for only those stars is decreased by a factor $f(t) = 20(1 + t/t_{end})$ to avoid runaway growth of massive sinks and to account for massive stars eventually running out of material that they can accrete. 
\par In the first case, $\alpha_i$ is drawn randomly from the same log-normal distribution at every time. This starting distribution $\alpha_0$ is centered at $10^{-6} M_{\odot}^{-1} \rm yr^{-1}$ with logarithmic spread $\sigma_{\log \alpha} = 0.75 $; these parameters are chosen to produce a reasonable range of values of $dM/dt$ ($10^{-11} - 10^{-5} M_{\odot} yr^{-1}$) given a typical range of masses, $0.01 - 25 M_{\odot}$. This configuration can build a power-law tail for the mass function, even though the range of $\alpha$ can span $\sim 4$ orders of magnitude. In the next case, in order to simulate an environment being depleted of accretable material, $\alpha_i$ is drawn from a distribution shifting to lower values over time where $\alpha (t) = \alpha_0 (1 - (t/t_{end})^2)$. The primary effect of this change is that, as expected, it takes a longer time for the power-law tail to develop. In the last case, in addition to the decreasing values of $\alpha_i$, the stars are formed not all at once, but throughout the simulation, i.e. stars are drawn from $\xi(M_0)$ at a specified star formation rate -  50 new stars are initialized every 0.125 Myr. The 
slope of accretion rates vs stellar mass in Figure \ref{fig:toybhl} is slightly steeper than mass squared, even though each simulation has the accretion rate explicitly prescribed to scale with $M^2$.For less massive stars, this 
departure from the $M^2$ relation is a result of the scatter in the $\alpha$ values and the shape of the mass function itself, rather than a difference in accretion behavior. At the most massive end, the mass loss rates begin to fall below the $M^2$ trend due to the additional factor $f(t)$ used when stars exceed a threshold mass in order to avoid runaway accretion, as described above. This reduction in accretion rates also 
results in the spike or pile up of objects at around
$\log M \sim 0.8$ in the mass functions.

While the above analysis indicates that BHL-type gravitationally-focused accretion can
produce upper-mass power-law mass functions, the emergence of
the power law is dependent upon a) having 
a mechanism for starting a ``seed'' mass from a
non-BHL process (via thermal fragmentation,
for example) and b) growing masses well beyond
the initial seed mass distribution.  This suggests that the star cluster mass function
should provide a better example of gravitational focusing producing a power law mass function \citep[e.g.,][]{kuznetsova17}, as the
thermal fragmentation scale relative to the final mass should be much smaller in the cluster case than for the stellar mass function.

\section{Summary}
 
We have presented isothermal hydrodynamic simulations with sink formation to explore the effects of gravity in accretion.  We found that, by including the mass in the near environment of sinks, the accretion rates exhibit an $M^2$ dependence similar to that in simple BHL accretion.  Our findings strongly suggest that the apparent inconsistencies found in previous analyses were the result of only considering the sink mass, which at early times strongly underestimates the relevant gravitating mass.
We also developed toy models to illustrate that several features seen in simulations that are incompatible with the simple Z82 model -- complex time dependence of mass addition, departure from the $M^2$ dependence of accretion rate on mass in snapshots, range of accretion timescales, correlation of final mass with time of initial formation -- are compatible with gravitationally-focused accretion.
In turn, this
 provides an explanation  for how the $\Gamma \rightarrow -1$ mass function tends to develop. 

The gravitationally-focused accretion outlined here
is a major feature of the ``competitive accretion''
model for star formation
\citep[e.g.,][]{bonnell01a,bonnell01b,bonnell08},
though in our case a global tidal potential plays no significant role. 
The slightly steeper Salpeter slope in our interpretation results from the exhaustion of material, especially as stellar feedback can be an important factor in dispersing gas; simulations with feedback are needed explore this conjecture.

\acknowledgments
This work was supported in part by NASA grant NNX16AB46G, by the University of Michigan. JBP acknowledges financial support from UNAM-PAPIIT grant number IN110816.
We used computational resources and services provided by Advanced Research Computing at the University of Michigan, Ann Arbor, and by the Information and Technology Services at UNC Chapel Hill.

\appendix
\subsection*{Resolution Comparison}

\renewcommand{\thefigure}{A\arabic{figure}}
\par The main results of this paper are largely unchanged by a factor of $2$ increase in resolution (dynamic range). Here we show figures corresponding to Figures \ref{fig:mdotm}  and \ref{fig:mf} but at a resolution of $1024^3$, such that the patch is now at a scale of 0.05 pc instead of 0.1 pc. The primary differences between the two runs is that at $1024^3$ more sinks are created and  the power law of the mass function tends to better developed within 2.2 Myr for the higher resolution runs. 

\par There is less mass in the patch gas as the physical size of the patch is now reduced, however, as the scale of structures tends to decrease while the density increases in an isothermal simulation, this difference does not appear to be very large when comparing the last panels of Figure \ref{fig:mf} and  Figure \ref{fig:1024mf}, both shown at t = 2.2 Myr ($\sim$ 1.3 $t_{\mathrm{ff}}$).

\par The reduction in physical accretion radius does not have an appreciable effect on the accretion rates and only a small effect on patch masses, still yielding the $dM/dt \propto M^2$ behavior seen in the fiducial run.

\begin{figure*}[h]
    \centering
    \includegraphics[width=0.9\paperwidth]{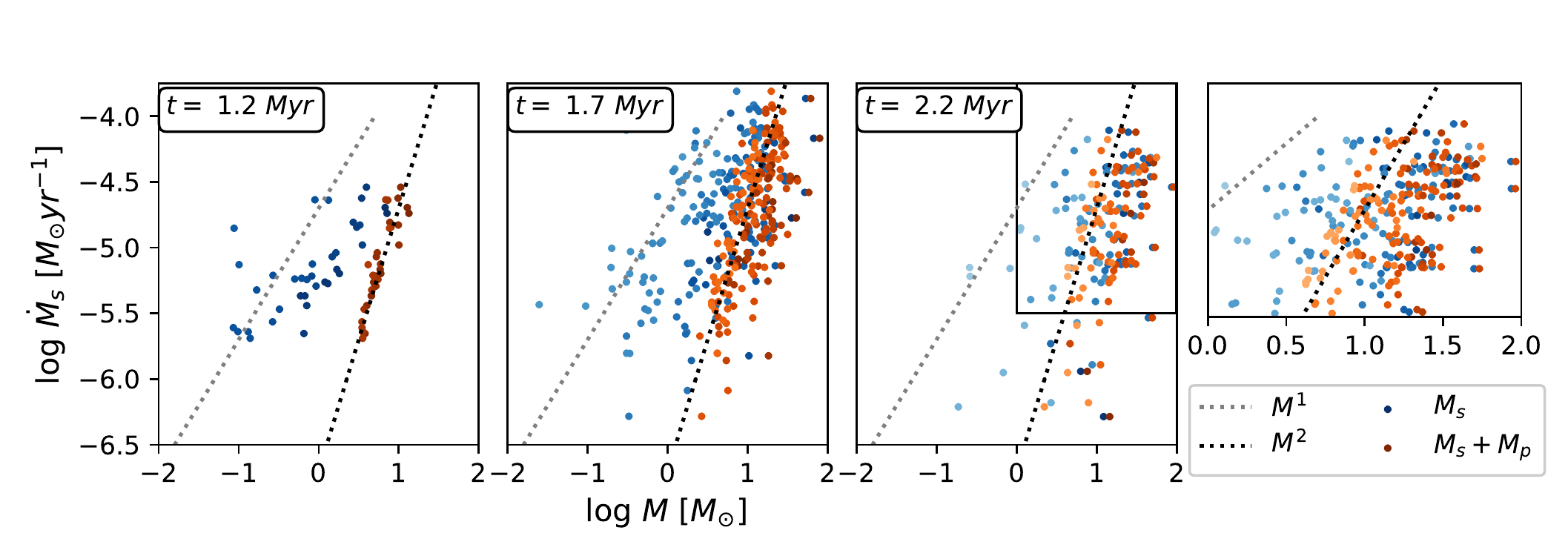}
    \caption{Sink accretion rate vs sink mass, $M_s$, and total enclosed patch mass, $M_s + M_p$ shown for t = 1.2, 1.7, and 2.2 Myr. The gray and black dotted lines have slopes of 1 and 2, respectively.The $\dot{M}_s^2 \propto (M_s + M_p)^2$ relationship still holds for $N_{cell} = 1024^3$, even when patch sizes are smaller.}
    \label{fig:1024mdot}
\end{figure*}

\begin{figure*}[h]
    \centering
    \includegraphics[width=0.9\textwidth]{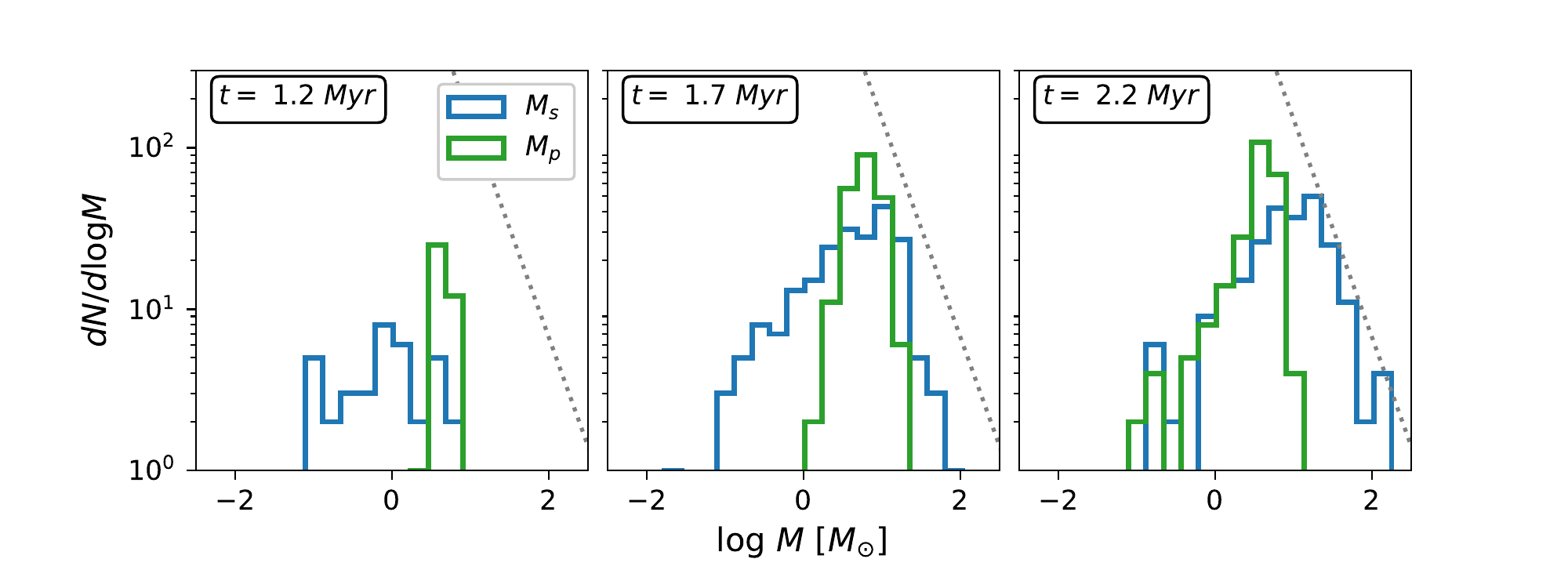}
    \caption{Mass function for the sink mass $M_s$ (blue) and the patch mass $M_p$ (green) at 1.2, 1.7, and 2.2~Myr, for a resolution of $1024^3$. The gray dotted line is the $\Gamma=1.35$ Salpeter slope. The patch gas masses are somewhat smaller due to the decreased patch size. }
    \label{fig:1024mf}
\end{figure*}

\newpage

\bibliographystyle{aasjournal}
\bibliography{refs1}

\end{document}